\documentclass[prb,aps,twocolumn]{revtex4}
\usepackage{graphicx}
\usepackage{dcolumn}
\usepackage{bm}

\begin{document}
\title {Electrochemical fabrication of ultra-low noise metallic
nanowires with hcp crystalline lattice}
\author{Amrita Singh\footnote[1]{electronic mail:amrita@physics.iisc.ernet.in}}
\author{T. Phanindra Sai}
\author{Arindam Ghosh}
\address{Department of Physics, Indian Institute of Science, Bangalore 560 012, India}


\begin{abstract}
We experimentally demonstrate that low-frequency electrical noise
in silver nanowires is heavily suppressed when the crystal
structure of the nanowires is hexagonal closed pack (hcp) rather
than face centered cubic (fcc). Using a low-potential
electrochemical method we have grown single crystalline silver
nanowires with hcp crystal structure, in which the noise at room
temperature is two to six orders of magnitude lower than that in
the conventional fcc nanowires of the same diameter. We suggest
that motion of dislocations is probably the primary source of
electrical noise in metallic nanowires which is strongly
diminished in hcp crystals.
\end{abstract}


\maketitle

Metallic nanowires are integral components of several nanoscale
electronic circuits, particularly in crossbar interconnect
architectures.~\cite{snider} Both experiments~\cite{Steinhogl,
bid3} and electrical modelling~\cite{Jhou} have often addressed
the effects of size on average electrical resistivity ($\rho$) of
the nanowires, but very little is known about the low-frequency
$1/f$-type electrical noise in these systems which can seriously
impede their application in nanoelectronics.~\cite{collins,Bid1}
Moreover, recent experiments suggest that the magnitude of noise
in nanowires can indeed be much larger than that generally
observed in thin polycrystalline metal films.~\cite{Aveek} Hence
our objective here is to address two issues: (1) what is the
microscopic origin of low-frequency electrical noise in nanowires,
and (2) can the noise in nanowires be suppressed or reduced by
appropriate engineering of growth and structural parameters.

The power spectral density $S_R$ of low-frequency $1/f$ noise in
resistance $R$ is normalized as, \vspace{-.25cm}
\begin{displaymath}
\label{eq1} S_R(f)=\frac{\gamma_HR^2}{N_e(f/Hz)^\alpha}
\vspace{-.2cm}
\end{displaymath}

\noindent where $N_e$ is the total number of electrons, and
$\gamma_H$ is the phenomenological Hooge parameter that depends on
the material properties, such as nature and kinetics of disorder,
scattering cross-section of electrons, crystallinity, and so
on.~\cite{Pelz,Hooge} In thin polycrystalline metallic films,
$\gamma_H$ generally lies in the range $\sim 10^{-3} - 10^{-5}$,
which is predominantly due to the migration of point defects along
the grain boundaries.~\cite{Scofield,Koch,weissman} Surprisingly
though, the value of $\gamma_H$, in electrochemically grown single
crystalline fcc silver nanowires (AgNWs) ($\approx 15$~nm diameter
) was found to be very large ($\gamma_H \sim 10^{-1} - 10^{-2}$),
even when the grain boundaries are expected to be absent. Although
an explanation based on Rayleigh-Platteau instability has been
proposed for narrow nanowires~\cite{Aveek}, the source of noise in
wires of larger diameter ($\sim 100$~nm) is not clearly
understood. Moreover, whether other defect kinetics, such as
thermally activated movement of dislocations,~\cite{Nishida}
contribute to the observed noise still remains uncertain.

The nature of crystallinity is known to have a profound influence
on the kinetics of dislocations.~\cite{Hull} Since it is
intimately connected to plasticity, a study of noise can also be
relevant to understand the intrinsic structural aspects of the
nanowires. Hence, to evaluate the role of defect kinetics on noise
in nanowires, we have carried out detailed electrical
characterization of single crystalline AgNWs in both fcc and hcp
forms. We find that the value of $\gamma_H$, at room temperature
($T \approx 300$~K) in hcp AgNWs, is several orders of magnitude
lower than that in fcc AgNWs, which indicates that dislocation
motion may be the dominant source of electrical noise in metallic
nanowires.

Although silver is fcc in bulk crystalline form, a competition
between the surface and internal packing energies has been shown
to stabilize the hcp structure in both nanoparticles~\cite{pushan}
and nanowires~\cite{Liu} below certain critical dimensions.
Conventional over-potential electrodeposition (OPD) technique,
which employs inter-electrode potential ($V_d$) larger than the
equilibrium Nernst potential ($E_0$) yields mainly fcc AgNWs, when
the template pore size is more than 50 nm. Recently, a modified
electrochemical route was suggested, where the growth takes place
at $|V_d| \ll |E_0|$, with the resulting AgNWs stabilized entirely
in the single crystalline hcp phase even for nanowire diameters as
large as $100$~nm.~\cite{Singh} Even though the precise growth
mechanism for the latter, called the low-potential
electrodeposition (LPED), is not well understood, the control on
the crystal structure of the nanowires allows us to investigate
electrical noise in nanowires of the same diameter and material,
but different dislocation kinetics.

We have used polycarbonate templates (Whatman International Ltd.)
with average pore size $\approx 60$ nm and pore length~$6$ $\mu$m,
which were coated with thin gold film on one side to serve as
anode. By injecting electrolyte through a metallic micro-capillary
(cathode) mounted on a vertical micropositioning stage, the AgNWs
were locally electrodeposited over an area of about 1~mm$^2$ of
the template under both LPED and OPD conditions. The selected area
electron diffraction (SAED) patterns from similarly prepared
nanowires, as shown in Fig.~1a and 1b, show the OPD AgNWs in
expected fcc structure, while the LPED AgNWS in hcp phase (without
any trace of fcc component, as confirmed by the X-ray diffraction
(XRD) studies~\cite{Singh}). The time of local electrodeposition
was optimized (20 mins) to allow a small overgrowth of mushroom
heads, so that the contact resistance was extremely low ($\lesssim
0.2$ ~Ohms), when connected with conducting silver epoxy. The
local electrodeposition was used to restrict the number of
nanowires, and to ensure that the net resistance and noise are not
too small to measure. Fig.~1c displays  Scanning Electron
Microscope image of a typical device without electrical contacts.
The schematic of the electrical characterization set-up, which was
used for both resistivity and noise measurements, is sketched in
Fig.~1d. For the noise measurements, the device was included as
one of the arms of a Wheatstone bridge, followed by amplification
of the error signal, digitization and digital signal processing,
which allows simultaneous measurement of background as well as
bias dependent sample noise.~\cite{Scofield2} The apparatus can
measure voltage power spectral densities as small as
$10^{-21}$V$^2$/Hz. The details of data acquisition and digital
signal processing are described elsewhere.~\cite{Ghosh}

In Fig.~2a we have summarized the result of conductance
measurements on more than 100 devices, grown at $V_d$ ranging over
nearly five decades. In the LPED regime, the devices were
generally stable for $|V_d| \gtrsim 10$~mV (yield $\sim 40 -
50$\%), but become progressively unstable as $V_d$ was reduced,
with the life time being as small as few hours. We did not find
any specific trend in conductance as a function of $V_d$, probably
due to the fluctuations in the number of nanowires from one device
to the other, or any marked difference in the value of conductance
for the nanowires grown in OPD and LPED conditions. To study the
behavior of noise in the LPED regime, we have focused on AgNW
device grown at $V_d = 100$~mV, although most of the other stable
devices also showed qualitatively similar behavior.

To evaluate $\gamma_H$ quantitatively, the number of nanowires in
the device under study was estimated using the method outlined in
Ref.[3]. We first fitted the $R-T$ data with Bloch-Gr\"{u}neisen
formula, which is shown in Fig.~2b and 2c for two typical OPD and
LPED devices, respectively. The Debye temperature ($\Theta_D$) was
extracted as a fit parameter, and found to be 240~K for OPD and
225~K for the LPED nanowires, which are about $10-20$\% higher
than that of the bulk value ($\Theta_{D0} = 204$~K). Following the
method described in Ref.[3], the number of nanowires
 in LPED and OPD devices were estimated as 28 and 140, respectively.

Fig.~3 illustrates the power-spectral density ($S_R$) in the LPED
(Fig.~3a) and OPD (Fig.~3b) devices over nearly three decades of
frequency ($f$) at different values of $T$, stabilized at better
than $\pm 10$~ppm. The background potential fluctuations (Johnson
noise, amplifier noise etc.) were below $10^{-19}$~V$^2$/Hz during
all the measurements. The noise power drops rapidly with
decreasing $T$ for fcc AgNWs, while it is nearly independent of
$T$ for hcp AgNWs. $\gamma_H$ can be weakly frequency dependent if
$S_R$ deviates from pure power law over the observed frequency
range, and for comparison, $\gamma_H$ measured at 1 Hz, was used.
The temperature dependence of $\gamma_H$ in Fig.~4a shows that at
the room temperature, the absolute value of $\gamma_H$ in fcc
AgNWs is three orders of magnitude higher than that of hcp AgNWs,
but becomes roughly equal at $\sim 100$~K. Measurements on a large
number of LPED AgNW devices show that the noise in these systems
at $T \approx 300$~K may vary with growth parameters, intrinsic
disorder and thermal cycles. Nevertheless, the value of $\gamma_H$
was always found to be much lower than that of the OPD devices,
and expressed as the bands corresponding to the growth potential
$V_d$ in Fig.~4d.

The contribution of Rayleigh instability to noise in our AgNWs of
diameter $\sim 60$~nm is expected to be negligibly small. Noise
from pipe diffusion of defects along a dislocation line has been
considered in single crystalline Aluminium,~\cite{Homberg} but in
such a case the LPED and OPD devices would probably display
similar noise magnitudes. However, if the noise arises from motion
of dislocations themselves then its suppression in LPED AgNWs can
be attributed to less number of slip systems (only basal plane) in
hcp crystal than fcc OPD systems.~\cite{Hull} The strong
$T$-dependence of noise in OPD AgNWs can be understood through a
thermally activated kinetics of dislocations under the influence
of internal stresses generated during growth~\cite{Kamel}.
However, the weak $T$-dependence of noise in hcp nanowires
indicates that activated kinetics, similar to that discussed by
Dutta, Dimon and Horn~\cite{Dutta}, are unlikely to be applicable
in this case. An understanding of this may be linked to the
high-energy crystal structure of LPED AgNWs which drives these
systems away from the thermal equilibrium. The motion of
dislocation would then correspond to transition between metastable
states that are separated by strong potential
barriers,~\cite{Chandni} and would occur only under the influence
of internal stresses, rather than temperature fluctuations.

The residual resistivity ratio ($R_{\rm300K}/R_{\rm4.2K}$) for
LPED AgNWs was generally much smaller than OPD devices (Figs.~2c
and 2d), which indicates the presence of additional disorders,
such as vacancy clusters, that can lock the dislocation motion by
forming Cottrell atmosphere around the dislocations. This may
result in further lowering of noise level in hcp
nanowire.~\cite{Hull} To substantiate this, we show the
temperature dependence of frequency exponent $\alpha$ for both
LPED and OPD systems in Figs.~4b and 4c respectively. The
magnitude of $\alpha$ is consistently slightly larger in the OPD
devices, and approaches $\sim 1.3 - 1.4$ for temperatures above
$200$~K. Note that for long-range diffusion of the scattering
species, $\alpha$ is expected to be 1.5, which has been
demonstrated for dislocation dynamics under the long-range stress
field during plastic deformation.~\cite{laurson}

In summary, we have demonstrated that the $1/f$-noise in metallic
nanowires is highly crystal structure dependent, and the noise
magnitude in hcp AgNWs fabricated by low potential
electrodeposition process is far less than that of conventional
fcc AgNWs. The lower noise magnitude in hcp nanowires could be
attributed to the dislocation dynamics restricted only in the
basal plane, and also the locking of dislocation motion by point
defects and high energy barriers.

\textbf{Acknowledgement} \linebreak We thank Ministry of
Communication and Information technology, Government of India, for
a funded project.

\newpage

\begin{figure}
\vspace{6cm}
\includegraphics[width=10cm,height=10cm]{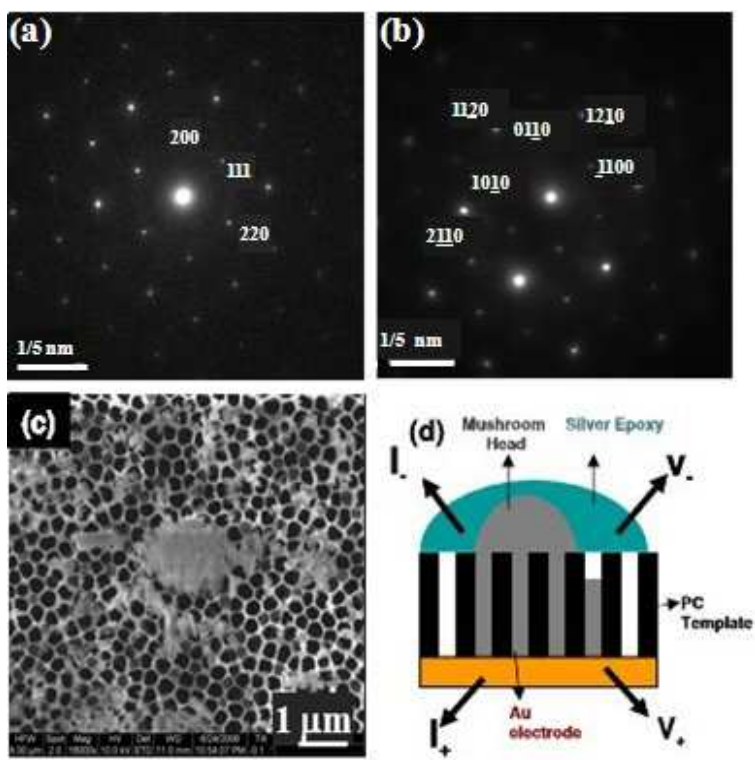}
\caption{ Selective area electron diffraction (SAED) pattern of
Silver nanowires (AgNWs) grown by (a) over potential
electrodeposition (OPD) process, (b) low potential
electrodeposition (LPED) process, (c) Scanning Electron Microscope
(SEM) image of AgNWs embedded in polycarbonate membrane showing
mushroom heads, (d) schematic of contacted AgNWs for electrical
characterization.} \label{figure1}
\end{figure}

\begin{figure}
\includegraphics[width=10cm,height=14cm]{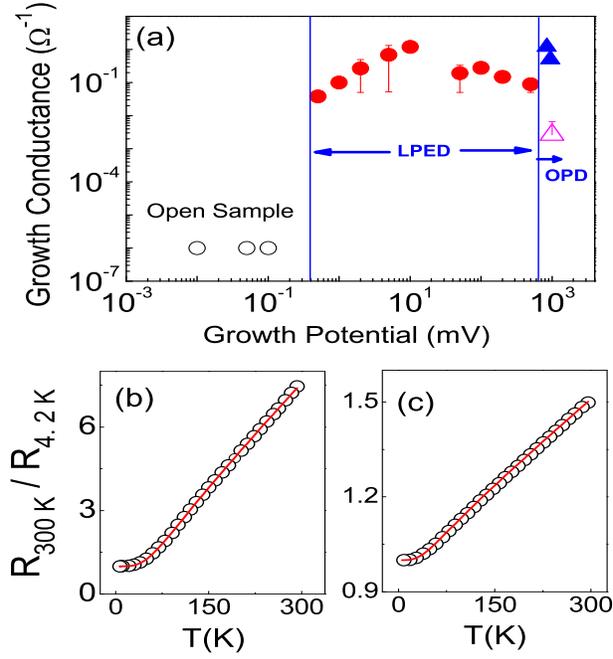}
\vspace{-5cm}
\caption{ (a) Growth-conductance vs growth-potential curve of as
grown silver nanowires (AgNWs). Filled triangles correspond to the
nanowires, grown by conventional OPD process and the open triangle
corresponds to the AgNWs, grown at 1000 mV with the  2M
electrolyte, (b) Resistance vs temperatute $(R-T)$ curve of face
centered cubic (fcc) AgNWs, (c) $R-T$ plot of Hexagonal closed
packed (hcp) AgNWs. The open circles represent the experimental
data while the solid line (red) shows the Bloch-Gr\"{u}neisen fit
to the experimental $R-T$ data.} 
\label{figure2}
\end{figure}

\begin{figure}
\includegraphics[width=15cm,height=8cm]{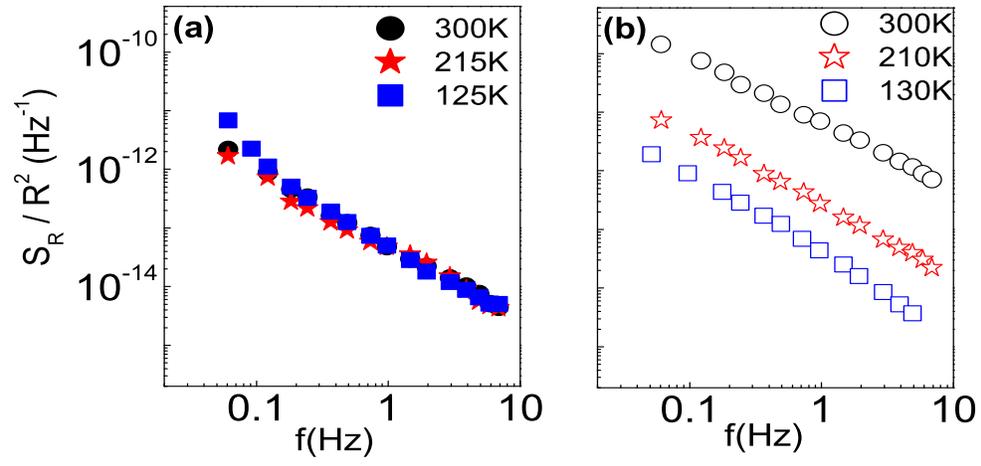}
\caption{ Normalized power spectra vs frequency plot of (a) hcp
silver nanowires (AgNWs) at temperatures $300$~K, $215$~K and
125~K, (b) fcc AgNWs at temperatures $300$~K, $210$~K and
$130$~K.} \label{figure3}
\end{figure}

\begin{figure}
\includegraphics[width=10cm,height=14cm]{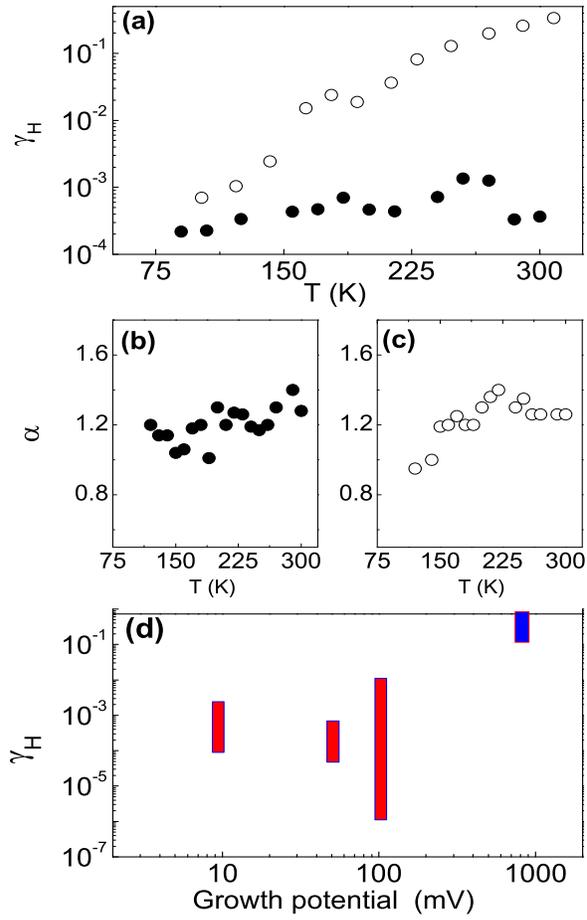}
\caption{ Variation of (a) Hooge parameter $\gamma_H$ ($f$ = 1 Hz)
with temperature for hcp nanowires (filled circle) and fcc
nanowires (open circle), (b) frequency exponent $\alpha$ with
temperature for hcp nanowires (filled circle) and fcc nanowires
(open circle), (c) Hooge parameter $\gamma_H$, in the form of
bands, with growth potential of as grown silver nanowires at room
temperature.} \label{figure4}
\end{figure}

\end{document}